\documentclass[11pt]{article}

\usepackage[margin=1in]{geometry}
\usepackage{amsmath, amssymb}
\usepackage{graphicx}
\usepackage[hidelinks]{hyperref}
\usepackage{bm}
\usepackage{tikz}
\usepackage{float}
\usetikzlibrary{arrows.meta, positioning}

\title{Karhunen--Lo\`eve Expansion--Based Residual Anomaly Map for Resource-Efficient Glioma MRI Segmentation}
\author{Anthony Joon Hur}
\date{\today}

\begin{document}

\maketitle

\begin{abstract}

Accurate segmentation of brain tumors is essential for clinical diagnosis and treatment planning. Deep learning is currently the state-of-the-art for brain tumor segmentation, yet it requires either large datasets or extensive computational resources that are inaccessible in most areas. This makes the problem increasingly difficult: state-of-the-art models use thousands of training cases and vast computational power, where performance drops sharply when either is limited.

The top performer in the Brats GLI 2023 competition relied on supercomputers trained on over 92,000 augmented MRI scans using an AMD EPYC 7402 CPU, six NVIDIA RTX 6000 GPUs (48GB VRAM each), and 1024GB of RAM over multiple weeks. 

To address this, the Karhunen--Lo\`eve Expansion (KLE) was implemented as a feature extraction step on downsampled, z-score normalized MRI volumes. Each 240$\times$240$\times$155 multi-modal scan is reduced to four $48^3$ channels and compressed into 32 KL coefficients. The resulting approximate reconstruction enables a residual-based anomaly map, which is upsampled and added as a fifth channel to a compact 3D U-Net.

All experiments were run on a consumer workstation (AMD Ryzen 5 7600X CPU, RTX 4060Ti (8GB VRAM), and 64GB RAM) while using far fewer training cases. This model achieves post-processed Dice scores of 0.929 (WT), 0.856 (TC), and 0.821 (ET), with HD95 distances of 2.93, 6.78, and 10.35 voxels. These results are significantly better than the winning BraTS 2023 methodology for HD95 distances and WT dice scores. This demonstrates that a KLE-based residual anomaly map can dramatically reduce computational cost and data requirements while retaining state-of-the-art performance.
\end{abstract}

\section{Background}

\subsection{Clinical Impact of Gliomas}
\subsubsection{Epidemiology and Clinical Impact}

Gliomas are the most common type of primary malignant brain tumor, accounting for approximately 75-80\% of all malignant primary brain tumors in adults, with glioblastoma being the most aggressive and prevalent subtype \cite{Ostrom2023}. In the United States, there are more than 24,000 new cases of malignant brain and other central nervous system tumors diagnosed annually, the majority of which are gliomas \cite{Ostrom2023}.

Patients with glioma suffer high mortality rates and limited long-term survival. Low-grade gliomas are generally slower growing and may be managed with surgical resection and adjuvant therapy; however, many progress to higher-grade malignancies over time \cite{Louis2021}. High-grade gliomas exhibit rapid infiltration into surrounding brain tissue, resistance to treatment, and recurrence. Glioblastoma has a median survival of approximately 12-15 months following diagnosis and a 5-year survival rate below 10\%. \cite{Louis2021,Ostrom2023}.

Gliomas all around the world impose a significant public health burden, as glioma patients often experience severe neurological deficits, including cognitive impairment, motor dysfunction, and seizures, which substantially reduce quality of life and functional independence \cite{WHO2022}. Given these outcomes, accurate diagnosis, treatment planning, and longitudinal monitoring are critical for improving patient survival and quality of care, motivating the development of precise imaging-based tumor delineation methods.

\subsubsection{Tumor Characterization and Segmentation}

Brain tumors, gliomas in this case, originate from glial cells that support nerve cells. This results in the uncontrolled division and growth of cells, leading to inflammation and death of brain cells. The NHS categorizes these brain tumors into four separate grades, with grades 1 and 2 classified as low-grade, non-cancerous, benign tumors, which are less likely to recur after treatment. On the contrary, grades 3 and 4 are categorized as high-grade, cancerous, malignant tumors that are more likely to spread throughout the brain and relapse after treatment. \cite{NHSbraintumors} Accurate representations of tumor subregions, including necrotic or non-enhancing tumor, peritumoral edema, and actively enhancing tumor, are essential for treatment planning, radiotherapy, and longitudinal monitoring, where multi-channel MRI volume scans are the standard for imaging modality. In the BraTS dataset, each patient has these four channels: T1-weighted (T1), contrast-enhanced T1 (T1Gd or T1c), T2-weighted (T2), and T2-FLAIR (FLAIR) volumes. 

Though, actual manual voxel-wise segmentation is time-consuming and prone to variability, requiring many hours to complete, for potentially biased results. By automating time-consuming tasks for radiologists, the use of AI in tumor segmentation drastically improves the efficiency of the treatments stated previously.

\begin{figure}[H]
    \centering
    % 5-panel PDF: T1, T1Gd, FLAIR, T2, T1Gd+label
    \includegraphics[width=0.95\textwidth]{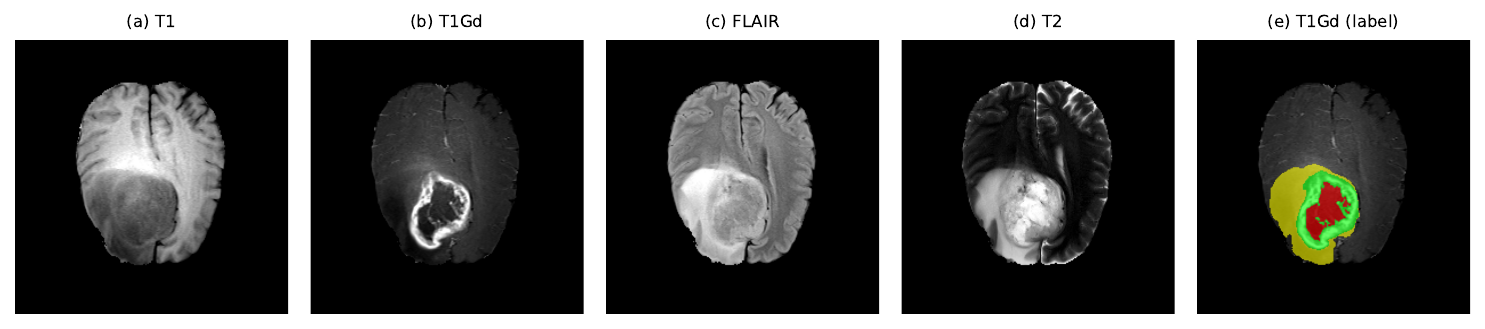}
    \caption{All four channels of a sample BraTS-GLI patient and the corresponding manual segmentation overlaid on the T1Gd image. From left to right: T1, T1Gd, FLAIR, T2, and T1Gd with label map (red = NCR/NET, yellow = ED, green = ET). Figure created by Anthony Joon Hur using Python.}
    \label{fig:modalities}
\end{figure}

\begin{figure}[H]
    \centering
    \includegraphics[width=0.95\textwidth]{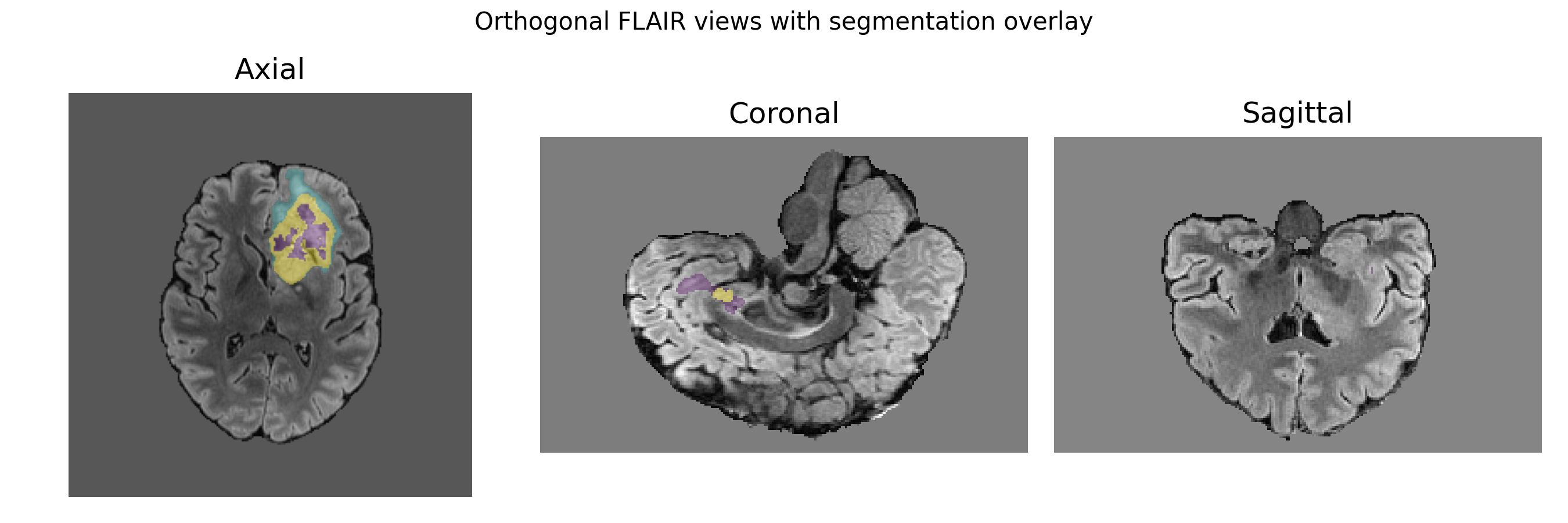}
    \caption{Orthogonal axial, coronal, and sagittal FLAIR views with segmentation overlay for a representative BraTS-GLI case. Figure created by Anthony Joon Hur using Python.}
    \label{fig:orthogonal}
\end{figure}

The BraTS challenge provides datasets and standardized evaluation metrics for automated tumor segmentation. Specifically, the 2023 adult glioma (GLI) challenge continues this by creating a competition to advance the field of automated brain tumor image analysis and segmentation by providing a benchmark for state-of-the-art algorithms with their datasets and results.

\subsection{Literature Review}

\subsubsection{Current Research 1: CNN-based BraTS segmentation}

While earlier approaches relied on features that fed into random forests and/or support vector machines, since around 2014-2017, deep convolutional neural networks (CNNs) have become the dominant approach to BraTS segmentation. Specifically, these tumor segmentation methodologies based on 3D U-Net and are now standard, in which the four MRI modalities are concatenated as channels, and the network performs voxel-wise classification into background. In the BraTS challenge, these modalities were T1-weighted (T1), T1-weighted with gadolinium contrast (T1Gd), T2-weighted (T2), and T2 Fluid Attenuated Inversion Recovery (FLAIR). This MRI data is represented as volumetric data; typically, a 4D tensor shape of C x D x H x W (where C = 4 due to the number of modalities). The dominant architecture used for BraTS is 3D U-Net, an extension of the 2D U-Net proposed by Ronneberger et al. (2015) \cite{ronneberger2015unet}, specifically adapted for volumetric segmentation by Çiçek et al. (2016) \cite{cicek}. 

This architecture follows an encoder-decoder structure, in which the encoder path consists of repeated blocks of 3D convolutions, batch normalization, and non-linear activators, then max-pooling operations. This allows for the path to capture global context and semantic information, at the cost of spatial resolution. In order to recover this spatial resolution, the decoder up-samples or transposes convolutions, allowing for precise boundaries of the tumor.

Additionally, feature maps from the encoder are concatenated with the corresponding layers in the decoder in a methodology called Skip connections. This allows the network to combine deep, semantic information with high-resolution surface details, mitigating the information loss caused by down-sampling.

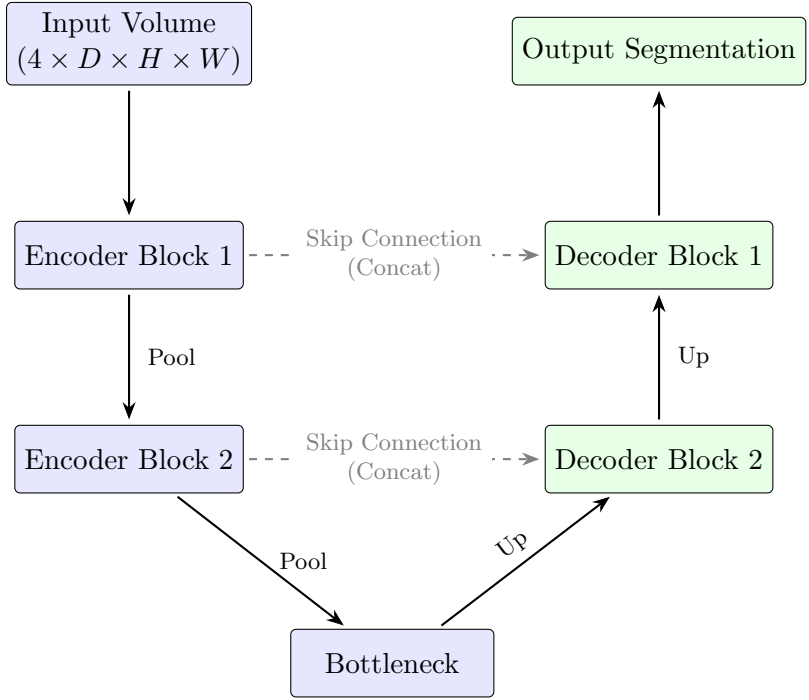
\begin{figure}[H]
    \centering
    \begin{tikzpicture}[
        node distance=1.8cm, % Adjusted vertical spacing
        layer/.style={
            draw,
            rectangle,
            minimum width=2.7cm,
            minimum height=0.9cm,
            align=center,
            fill=white,
            rounded corners=2pt
        },
        arrow/.style={-Stealth, thick, shorten >=2pt, shorten <=2pt},
        skip/.style={-Stealth, dashed, thick, gray, shorten >=2pt, shorten <=2pt}
    ]

    % Encoder (Contracting Path)
    \node[layer, fill=blue!10] (input) {Input Volume\\($4 \times D \times H \times W$)};
    \node[layer, below=of input, fill=blue!10] (enc1) {Encoder Block 1};
    \node[layer, below=of enc1, fill=blue!10] (enc2) {Encoder Block 2};
    \node[layer, below=of enc2, xshift=3.5cm, fill=blue!10] (bottleneck){Bottleneck};

    % Decoder (Expanding Path)
    % FIX: Align decoders relative to their ENCODER counterparts
    % This ensures the skip connections are horizontal straight lines
    \node[layer, right=4cm of enc2, fill=green!10] (dec2) {Decoder Block 2};
    \node[layer, right=4cm of enc1, fill=green!10] (dec1) {Decoder Block 1};
    \node[layer, above=of dec1, fill=green!10] (output) {Output Segmentation};

    % Connections (Main Path)
    \draw[arrow] (input) -- (enc1);
    \draw[arrow] (enc1) -- node[midway, right=3pt, font=\footnotesize] {Pool} (enc2);
    \draw[arrow] (enc2) -- node[midway, right=3pt, font=\footnotesize] {Pool} (bottleneck);
    
    % FIX: The "Up" arrow now goes diagonally or orthogonally from Bottleneck to Dec2
    \draw[arrow] (bottleneck) -- node[midway, sloped, above, font=\footnotesize] {Up} (dec2);
    \draw[arrow] (dec2) -- node[midway, right=3pt, font=\footnotesize] {Up} (dec1);
    \draw[arrow] (dec1) -- (output);

    % Skip Connections
    % FIX: These are now straight horizontal lines, preventing overlap
    
    % enc2 -> dec2
    \draw[skip] (enc2) -- node[midway, fill=white, font=\footnotesize, align=center] {Skip Connection\\(Concat)} (dec2);

    % enc1 -> dec1
    \draw[skip] (enc1) -- node[midway, fill=white, font=\footnotesize, align=center] {Skip Connection\\(Concat)} (dec1);

    \end{tikzpicture}
    \caption{Schematic representation of the 3D U-Net architecture used for BraTS segmentation. Dotted lines represent skip connections that concatenate encoder feature maps with decoder layers. Figure created by Anthony Joon Hur using LaTeX.}
    \label{fig:unet_arch}
\end{figure}

\subsubsection{Current Research 2: nnU-Net}

More recent BraTS-winning solutions build on nnU-Net, a self-configuring U-Net framework that automatically adapts network topology and hyperparameters to the dataset fingerprint. The BraTS 2023 GLI winning solution uses nnU-Net as a base and then expands it with synthetic tumor generation and intensity-based augmentations, before grouping several methodologies (such as Swin UNETR and other previous winning models) across folds. Training and inference are performed on an IKIM cluster node with six NVIDIA RTX 6000 GPUs (48 GB VRAM each), 1024 GB RAM, and an AMD EPYC 7402 24-core CPU. This infrastructure allows the authors to train and ensemble dozens of models, but it also raises the barrier for smaller labs as it requires increasingly more time and computational power to compute. \cite{ferreira2024brats}.

\subsubsection{Current Research 3: Optimization and Loss Functions}
Standard voxel-wise Cross-Entropy loss often fails in tumor segmentation due to severe class imbalance, in which the healthy tissue voxels completely outnumber the tumor voxels. To address this, current research prioritizes overlap-based loss functions, primarily the Dice Loss methodology \cite{milletari} or the Generalized Dice Loss methodology \cite{sudre}

\begin{equation*}
    \mathcal{L}_{Dice} = 1 - \frac{2 \bm{p} \cdot \bm{g} + \epsilon}{\|\bm{p}\|^2 + \|\bm{g}\|^2 + \epsilon}
\end{equation*}

$p_i$ is described as the predicted probability and $g_i$ is the ground truth label for voxel $i$. This loss directly optimizes the Dice Similarity Coefficient (DSC), the primary evaluation metric for the BraTS challenge. Many state-of-the-art implementations use a compound loss, optimizing a weighted sum of Cross-Entropy and Dice Loss to stabilize training.

\subsubsection{Current Research 4: KLE and anomaly-based data reduction}

A different research direction focuses on modeling nominal spatio-temporal fields and detecting anomalies as deviations from a low-rank structure. The deFOREST framework, for example, uses a discrete Karhunen--Lo\`eve expansion on multispectral satellite imagery to model typical vegetation and land cover patterns. A discrete KL basis provides an optimal low-dimensional approximation in the $L^2$ sense, and residuals from this approximation can be interpreted as anomaly scores. The associated supplement rigorously derives the discrete KL expansion, its optimality property, and concentration bounds for the residual components. 
\cite{deforest}

In this project, I adapt this discrete KLE machinery to 3D multi-modal brain MRI to create a residual anomaly map that helps make glioma segmentation feasible on modest hardware, implementing a residual anomaly map.

\subsection{Research Question}
\medskip
\noindent
\textbf{Research Question:} 
\emph{Can a KLE-based residual anomaly map allow a 3D U-Net trained on a consumer GPU to achieve glioma segmentation performance comparable to methods trained for weeks on multi-GPU clusters?}

\medskip

\subsection{Significance}

\subsubsection{Academic Significance}

From an academic standpoint, this work connects discrete KL expansions of random vectors with modern medical image segmentation. This project integrates KLE directly into the methodology to create a residual anomaly map, which is then input into a standard 3D U-Net. The project demonstrates how low-rank $L^2$ approximations can be used to reduce the effective dimensionality of a large 3D segmentation task and how this interacts with deep learning.

\subsubsection{Technological Significance}

Many institutions do not have access to servers or supercomputers with extensive computational power. Showing that competitive BraTS performance can be achieved on a desktop with a Ryzen 5 7600X and a single RTX 4060 Ti lowers the barrier for hospitals, smaller labs, and schools to experiment with advanced segmentation pipelines. If KLE-based creation of a residual anomaly map can consistently reduce memory and compute while preserving accuracy, the same idea could be applied to other 3D medical imaging tasks, such as liver, lung, or cardiac segmentation.

Additionally, this methodology proves that segmentation models can be created with increasingly less data, allowing for segmentation models to be created for rarer kinds of tumors that may not have as much publicly available data.

\subsection{Research Hypothesis}

\medskip
\noindent
\textbf{Hypothesis.} 
\emph{A 3D U-Net that incorporates a KLE-based anomaly channel, constructed from a highly compressed KL representation of the dataset, can achieve Dice and HD95 metrics on BraTS 2023 GLI that are competitive with state-of-the-art cluster-based methods, despite using only a single consumer GPU with 8 GB VRAM.}

\medskip

The hypothesis is that the discrete KL basis captures the main modes of normal brain variation, while the reconstructed residual map highlights voxels whose intensity patterns cannot be explained by this low-rank model. By providing both the raw MRI and this compact anomaly signal, the network can rely on a 5th channel summary of “what is unusual,” rather than discovering everything from scratch from raw intensities.

\section{Materials and Methodology}

\subsection{Prior Models}

Most BraTS segmentation pipelines are built around 3D U-Net architectures or nnU-Net variants. They take as input four co-registered MRI modalities and produce voxel-wise class probabilities for background and three tumor-related labels. State-of-the-art methods rely on large 3D patches, extensive data, data augmentation, post-processing, and model ensembling. Although these methodologies may have better performance gradually, these models increase memory usage, training time, and implementation complexity, and they assume that users replicating these methodologies have access to the same, high-end cluster GPU hardware.

\subsection{Prior Mathematical Analysis}

Demonstrated below is the discrete Karhunen--Lo\`eve expansion. Let $(\Omega,\mathcal{F},\mathbb{P})$ be a probability space and consider a random vector
\[
    \bm{v}(\omega) = [v_1(\omega),\dots,v_n(\omega)]^\top \in L^2(\Omega;\mathbb{R}^n),
\]
with the covariance matrix
\[
    C = \mathbb{E}\big[(\bm{v} - \mathbb{E}[\bm{v}])(\bm{v} - \mathbb{E}[\bm{v}])^\top\big].
\]
Assuming $C$ is positive definite, there exists an orthonormal eigenbasis $\{\bm{\phi}_k\}_{k=1}^n$ with eigenvalues $\lambda_1 \ge \dots \ge \lambda_n > 0$ satisfying

\begin{equation*}
    C \bm{\phi}_k = \lambda_k \bm{\phi}_k .
\end{equation*}
\noindent
The discrete KLE states that there are zero-mean, unit-variance, mutually uncorrelated random variables $Y_k(\omega)$ such that
\noindent

\begin{equation*}
    \bm{v}(\omega)
    = \mathbb{E}[\bm{v}] + \sum_{k=1}^{n} \sqrt{\lambda_k}\, \bm{\phi}_k\, Y_k(\omega),
    \label{eq:kl-full}
\end{equation*}
with $\mathbb{E}[Y_k Y_\ell] = \delta_{k\ell}$. Truncating the series at $m < n$ terms gives
\begin{equation*}
    \bm{v}_m(\omega)
    = \mathbb{E}[\bm{v}] + \sum_{k=1}^{m} \sqrt{\lambda_k}\, \bm{\phi}_k\, Y_k(\omega).
    \label{eq:kl-trunc}
\end{equation*}
This discrete expansion is the best $m$-dimensional linear approximation in mean-squared error:
\[
    \mathbb{E}\big[\|\bm{v}(\omega) - \bm{v}_m(\omega)\|_2^2\big]
    = \sum_{k=m+1}^{n} \lambda_k
    \le \mathbb{E}\big[\|\bm{v}(\omega) - Q_m \bm{v}(\omega)\|_2^2\big],
\]
for any orthogonal projector $Q_m$ onto an $m$-dimensional subspace.

Defining the residual $\bm{r}(\omega) = \bm{v}(\omega) - \bm{v}_m(\omega)$, one can derive bounds on the probability that a given component $r[i]$ is large in terms of the eigenvalues and eigenvectors. In this case, large normalized residuals indicate anomalies relative to the nominal low-rank model.

In my implementation, $n = 4 \times 48 \times 48 \times 48 = 442{,}368$ and $m=32$. Thus each downsampled 4-channel brain volume is compressed to 32 coefficients, giving a compression ratio of approximately $13{,}824\times$.

\subsection{Methodology}

I use the BraTS 2023 Adult Glioma (GLI) Task 1 dataset, which builds on the BraTS benchmark series and the underlying TCGA-GBM and TCGA-LGG collections \cite{menze2015brats,baid2021rsnabrats,bakas2017advancing,bakas2017gbm,bakas2017lgg}. Each case consists of four co-registered MRI modalities (T1, T1Gd/T1c, T2, FLAIR) at $1$\,mm$^3$ resolution and a segmentation mask with labels $\{0,1,2,4\}$ for background, necrotic or non-enhancing tumor, edema, and enhancing tumor, respectively. Figure~\ref{fig:modalities} illustrates one example BraTS-GLI case and highlights the visual differences between modalities and the corresponding label map.

For each case, the four modalities are stacked into a tensor of shape $(4,H,W,D)$. Then, a Z-score normalization is applied over non-zero voxels for each channel, in which the mean and standard deviation are computed for each modality over voxels with intensity greater than zero. Each voxel is standardized using these statistics, where the background remains at zero after normalization.

All experiments are conducted on a single desktop machine equipped with an AMD Ryzen 5 7600X CPU, an NVIDIA RTX 4060 Ti GPU with 8 GB of VRAM, and 64 GB of system RAM. This configuration is deliberately modest compared to the IKIM cluster node used by the BraTS 2023 winning team, which offered six RTX 6000 GPUs with 48 GB VRAM each, 1024 GB of RAM, and a 24-core AMD EPYC CPU \cite{ferreira2024brats}. The goal of this project is to show that, with a KLE-based data residual anomaly map, competitive segmentation quality is still possible under these constrained resources.

KLE fitting is performed once, offline, on the CPU. For a subset of training cases, each normalized 4-channel volume is downsampled to a target spatial size of $48\times 48\times 48$ using trilinear interpolation. The resulting tensor of shape $(4,48,48,48)$ is flattened into a vector in $\mathbb{R}^{442{,}368}$, and stacking these vectors forms a data matrix $X$. A PCA model with $m=32$ components is then fit to $X$, producing a mean vector, a loading matrix whose columns correspond to the discrete eigenvectors, and eigenvalues. To estimate residual variance, each row of $X$ is reconstructed from its 32-dimensional representation, the residual is computed, and the variance of each feature is estimated across the training cases.

For any new volume, including those used during training and validation, the normalized 4-channel volume is downsampled and flattened to a vector $\bm{v}$. After subtracting the mean to obtain $\bm{v}_\text{c}$, the 32 KL coefficients are computed as $\bm{y} = W^\top \bm{v}_\text{c}$, and the low-rank approximation is reconstructed as $\hat{\bm{v}} = \bm{\mu} + W \bm{y}$. The residual $\bm{r} = \bm{v} - \hat{\bm{v}}$ is normalized by the estimated residual standard deviation to obtain a vector $\bm{z}$. This vector is reshaped back to $(4,48,48,48)$, averaged over channels to produce a single-channel low-resolution anomaly volume, and upsampled back to $(H,W,D)$ using trilinear interpolation. The anomaly map is then added on to the four MRI modalities, resulting in a 5-channel input volume for the segmentation network.

Due to VRAM technical limitations, training is performed on 3D patches of size $96\times 96\times 96$. Patches are sampled in a tumor-aware fashion: locations are drawn within the brain, and a patch is accepted only if the fraction of tumor voxels exceeds a small threshold; otherwise, a new location is sampled. This strategy reduces the number of nearly empty patches dominated by background.

The total of 1,251 labeled BraTS-GLI cases is randomly split into 1,064 cases for training and 187 cases for validation, which corresponds to approximately 15\% of the data being held out for validation. The segmentation model is a standard 3D U-Net. The encoder consists of repeated blocks with $3\times 3\times 3$ convolutions, instance normalization, and ReLU activation, followed by $2\times 2\times 2$ max pooling. The decoder mirrors this structure with transposed convolutions for upsampling and skip connections that concatenate the corresponding encoder feature maps. The input to the network has 5 channels (the four normalized MRI modalities plus the KLE anomaly map), and the final layer is a $1\times 1\times 1$ convolution that produces four logits per voxel.

The network is trained for 100 epochs using the Adam optimizer with a learning rate of $10^{-4}$ and batch size 2. The loss function is the sum of multi-class cross-entropy and soft Dice loss over the non-background classes. A ReduceLROnPlateau scheduler monitors the validation Dice score and reduces the learning rate when progress stalls.

\begin{figure}[H]
    \centering
    \resizebox{0.9\textwidth}{!}{%
    \begin{tikzpicture}[node distance=2.0cm, >=Latex]
        \tikzstyle{block} = [
            rectangle,
            draw,
            rounded corners,
            minimum width=3cm,
            minimum height=1cm,
            align=center
        ]

        % Top row
        \node[block] (input) {4-channel MRI\\(T1, T1Gd, T2, FLAIR)};
        \node[block, right=of input] (pre) {Normalization\\Downsampling};
        \node[block, right=of pre] (kle) {KLE (PCA)\\Compression \&\\Reconstruction};
        \node[block, right=of kle] (anom) {Residual-based\\Anomaly Map};

        % Bottom row: U-Net directly under anomaly map
        \node[block, below=1.9cm of anom] (unet) {3D U-Net\\(5-channel input)};
        \node[block, right=of unet] (seg) {Tumor\\Segmentation};

        % Arrows along top
        \draw[->] (input) -- (pre);
        \draw[->] (pre) -- (kle);
        \draw[->] (kle) -- (anom);

        % Inputs to U-Net
        \draw[->] (input.south) |- (unet.west);   % original 4 channels
        \draw[->] (anom.south) -- (unet.north);   % anomaly map

        % Output
        \draw[->] (unet) -- (seg);
    \end{tikzpicture}%
    }
    \caption{Flowchart of the proposed pipeline. The four MRI modalities are normalized and downsampled for KLE fitting, which produces a low-rank approximation and a residual-based anomaly map. The upsampled anomaly map is concatenated with the original modalities to form a 5-channel input for a 3D U-Net, which outputs the tumor segmentation. Figure created by Anthony Joon Hur using LaTeX.}
    \label{fig:pipeline}
\end{figure}
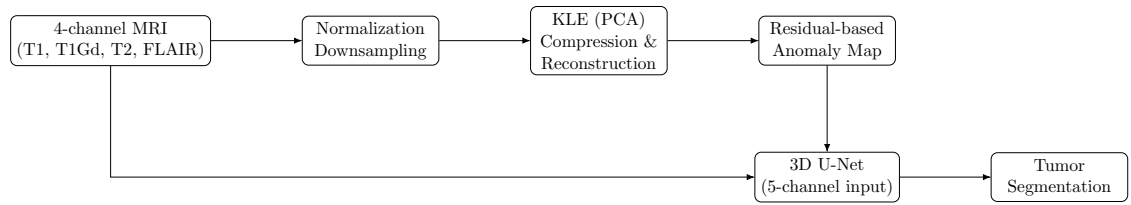

\section{Results}

Segmentation performance is evaluated on the 187-case validation set using the standard BraTS regions: whole tumor (WT), tumor core (TC), and enhancing tumor (ET). WT is defined as the union of labels $\{1,2,3\}$, TC as the union of labels $\{1,3\}$, and ET as label $\{3\}$ alone. For each region, I computed the Dice coefficient and the 95th percentile Hausdorff distance (HD95) in voxels.

After a simple post-processing step that removes small isolated components, the KLE-augmented 3D U-Net achieves mean Dice scores of approximately $0.929$, $0.856$, and $0.821$ for WT, TC, and ET, respectively. The corresponding HD95 values are about $2.93$, $6.78$, and $10.35$ voxels. The WT Dice is slightly higher than that reported by the BraTS 2023 GLI winning solution, while the TC and ET Dice are lower, likely due to the absence of ensembles and extensive synthetic augmentation. However, the HD95 values are consistently and substantially lower than those of the winning method across all three regions, indicating improved worst-case boundary localization and fewer extreme outliers.

\begin{figure}[H]
    \centering
    % 3x3 grid PDF: FLAIR / GT / prediction for three cases
    \includegraphics[width=0.9\textwidth]{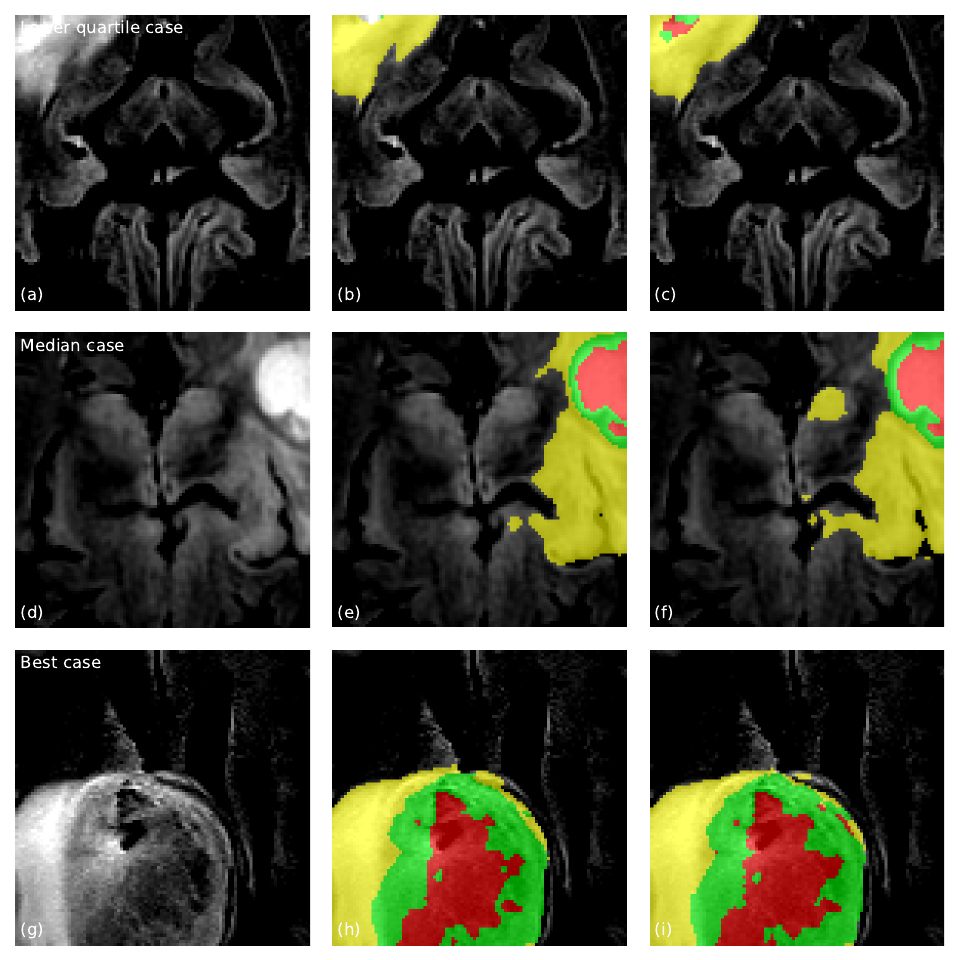}
    \caption{Qualitative validation examples segmented by the proposed KLE-augmented 3D U-Net. Rows correspond to a lower-quartile case, a median case, and the best case according to the mean WT/TC/ET Dice on the validation crop. Columns display the FLAIR slice, the FLAIR slice with ground-truth labels, and the FLAIR slice with the predicted segmentation. Red = NCR/NET, yellow = ED, green = ET. Figure created by Anthony Joon Hur using Python.}
    \label{fig:qualitative}
\end{figure}

Figure~\ref{fig:qualitative} shows three representative validation cases selected from the 187-case split. Each row corresponds to one case, ordered from a lower-quartile case (top row) to a median case (middle row) and a high-performing case (bottom row) according to their mean WT/TC/ET Dice on the validation crop. The columns display the FLAIR slice, the FLAIR slice overlaid with the ground-truth label map, and the FLAIR slice overlaid with the predicted segmentation. Red corresponds to NCR/NET, yellow to ED, and green to ET. These examples highlight that the KLE-augmented model usually captures the global tumor extent well, even for relatively difficult cases.

\begin{figure}[H]
    \centering
    \includegraphics[width=0.95\textwidth]{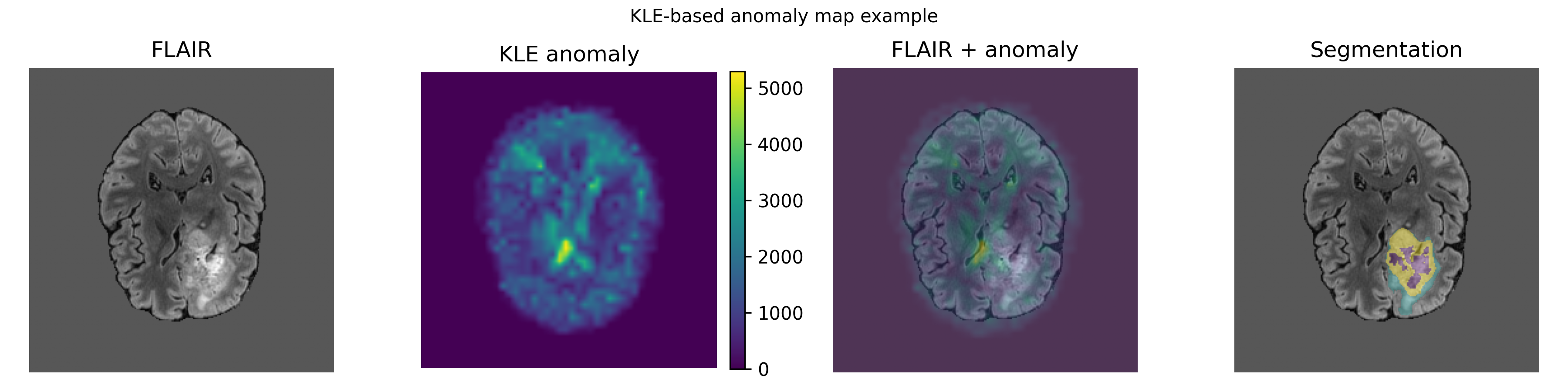}
    \caption{Example of the KLE-based anomaly map for a BraTS-GLI case. From left to right: FLAIR slice, normalized KLE anomaly map, anomaly map overlaid on the FLAIR image, and ground-truth tumor segmentation. High residual values concentrate in the tumor and peritumoral regions, indicating that the KLE residual provides a meaningful anomaly signal. Figure created by Anthony Joon Hur using Python.}
    \label{fig:kle_anomaly}
\end{figure}

Figure~\ref{fig:kle_anomaly} links the mathematical construction of the KLE residual to its practical effect on the images: regions that cannot be explained by the low-rank KL model appear as bright anomalies and align closely with tumor and edema. This visualizes how the anomaly channel emphasizes abnormal tissue and helps explain why providing it as a fifth input channel can improve boundary localization and robustness compared with using only the raw MRI intensities.

\begin{figure}[H]
    \centering
    \includegraphics[width=0.95\textwidth]{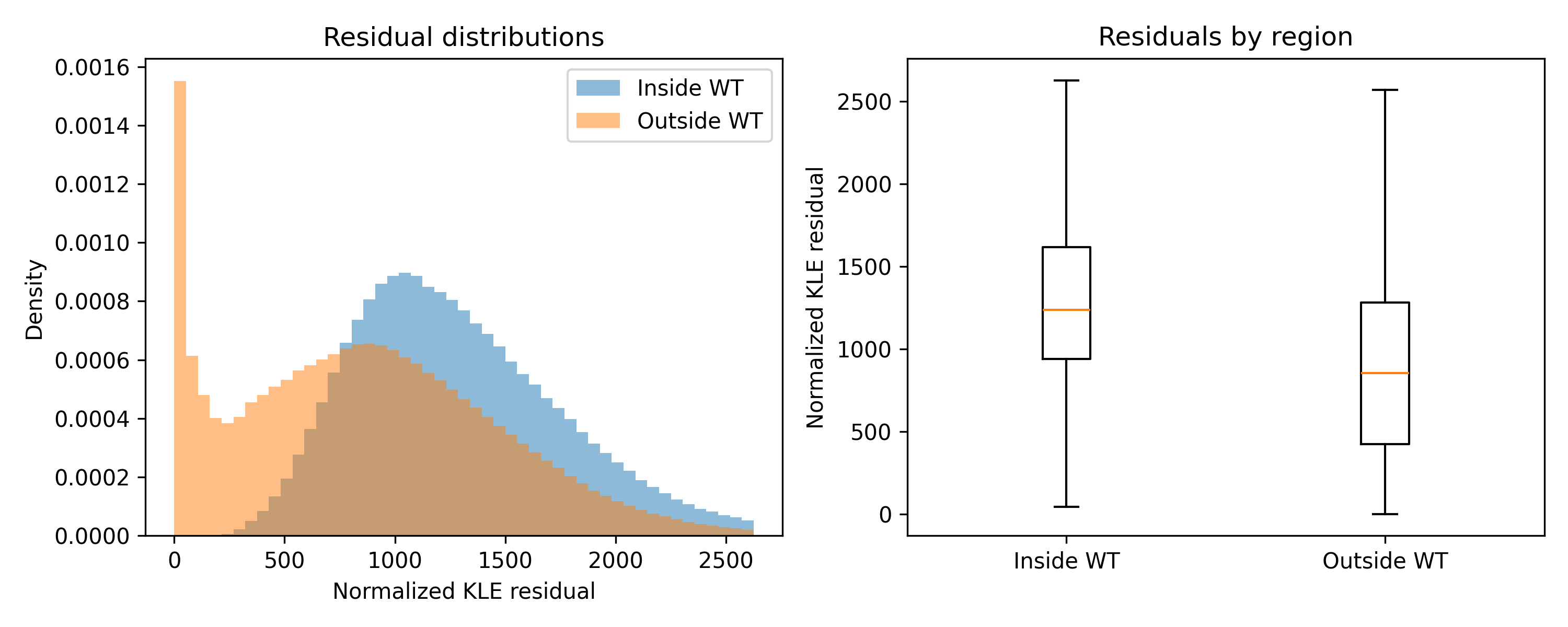}
    \caption{Distribution of normalized KLE residuals inside and outside the whole tumor (WT) region over the validation subset used for KLE fitting. Voxels within the tumor exhibit substantially higher residual values than voxels outside the tumor, further supporting the interpretation of the KLE residual as an anomaly measure. Figure created by Anthony Joon Hur using Python.}
    \label{fig:residuals}
\end{figure}

\section{Discussion}

\subsection{Data Analysis and Significance}

The central finding of this project is that a KLE-augmented 3D U-Net, trained on a single RTX 4060 Ti with 8 GB VRAM, can achieve WT Dice significantly higher than the BraTS 2023 winning solution and markedly better HD95 values across all regions, despite using far fewer computational resources. The cost, however, is slightly lower TC and ET Dice values.

From a data perspective, the KL expansion compresses each downsampled 4-channel volume from 442{,}368 features to just 32 coefficients, capturing the dominant modes of normal brain variation. The normalized residual is a compact, statistically meaningful anomaly descriptor that highlights tumor and edema regions. By providing this anomaly map as a fifth input channel, the network has access to information that would otherwise require fitting a very large model directly to high-dimensional raw data.

\subsection{Comparison of Research}

Compared with the BraTS 2023 GLI winning method, the proposed methodology is much simpler and requires increasingly less computational power. The winning pipeline uses three different architectures, three data augmentation strategies, five-fold cross-validation, and ensembling of 45 models on a multi-GPU cluster. Here, a single 3D U-Net with a KLE-based anomaly channel is trained once on a single consumer GPU. Despite this, the WT Dice is competitive, and the HD95 metrics are superior, at least on the internal validation split used in this project. In this sense, KLE plays the role of a data-efficient prior: instead of relying purely on more models and more GPUs, the method injects a low-rank statistical model of normal brain anatomy and focuses the network’s capacity on explaining the deviations from that model.

\subsection{Limitations}

This work has several limitations. All experiments are conducted on my own train/validation split of the BraTS 2023 GLI training set, not on the hidden test set used by the challenge organizers, which means that the comparison to the official leaderboard is indirect. The KL basis is learned from volumes that contain tumors; it does not attempt to separate healthy tissue when fitting the nominal model. Additionally, hyperparameters such as the number of KL components, the downsampled size, and the patch size were chosen based on memory constraints and a few exploratory runs, not on a systematic, justified basis.

\subsection{Future Research}

Future research could be applied to learn the KL basis from healthy-brain datasets or healthy-appearing tissue only, to obtain a purer nominal model, and then to compare it with the mixed tumor/healthy basis used here. Finally, broader generalization experiments on external clinical datasets and different scanners would be necessary to unto obtain a purer nominal model, and then understand how robust KLE-based data is in real-world conditions.

\section*{Conclusion}

This project demonstrates that the KLE-based data residual anomaly map can be used to achieve similar performance on the BraTS 2023 GLI dataset, utilizing limited computational power and reduced training data. By compressing each MRI volume into a small set of KL coefficients and utilizing the normalized reconstruction residual as an anomaly channel, a single 3D U-Net achieves competitive Dice scores and significantly improves HD95 compared to the BraTS 2023 winning methodology, while using less computational power. These results suggest that grounded data reduction techniques like the discrete KL expansion can be practical tools for resource-efficient deep learning in medical imaging.


\begin{thebibliography}{99}

\bibitem{ferreira2024brats}
A.~Ferreira, J.~A.~Reis, A.~M.~Castro, A.~C.~F.~Silva, and D.~M.~Martins, 
``How We Won BraTS 2023 Adult Glioma Challenge? Just Faking It! Enhanced Synthetic Data Augmentation and Model Ensemble for Brain tumor Segmentation,'' 
https://doi.org/10.48550/arXiv.2402.17317, 2024.

\bibitem{menze2015brats}
B.~H.~Menze, A.~Jakab, S.~Bauer, J.~Kalpathy-Cramer, K.~Farahani, J.~Kirby, \emph{et al.},
``The Multimodal Brain Tumor Image Segmentation Benchmark (BRATS),''
\emph{IEEE Transactions on Medical Imaging}, vol.~34, no.~10, pp.~1993--2024, 2015.
https://doi.org/10.1109/TMI.2014.2377694


\bibitem{milletari}
F. Milletari, N. Navab and S. -A. Ahmadi, "V-Net: Fully Convolutional Neural Networks for Volumetric Medical Image Segmentation," 2016 Fourth International Conference on 3D Vision (3DV), Stanford, CA, USA, 2016, pp. 565-571, https://doi.org/10.48550/arXiv.2510.14092

\bibitem{deforest}
J.~E.~Castrill\'on-Cand\'as, H. Gu, C. Meredith, Y. Li, X. Tang, P. Olofsson, M. Kon,
``deFOREST: Fusing Optical and Radar Satellite Data for Enhanced Sensing of Tree-loss,''
\emph{IEEE Transactions on Geoscience and Remote Sensing}, 2025, https://doi.org/10.48550/arXiv.2510.14092

\bibitem{Louis2021}
Louis, D. N., Perry, A., Wesseling, P., Brat, D. J., Cree, I. A., Figarella-Branger, D., \dots Ellison, D. W. (2021).
The 2021 WHO classification of tumors of the central nervous system: A summary.
\textit{Acta Neuropathologica, 142}(4), 493--516.
https://doi.org/10.1007/s00401-021-02385-4

\bibitem{NHSbraintumors}
NHS, "Brain tumours," NHS.UK, 2023, https://www.nhs.uk/conditions/brain-tumours/

\bibitem{cicek}
\"O.~\c{C}i\c{c}ek, A. Abdulkadir, S.~S. Lienkamp, T. Brox, and O.~Ronneberger.
(2016). 3D U-Net: Learning Dense Volumetric Segmentation from Sparse Annotation.
https://doi.org/10.48550/arxiv.1606.06650

\bibitem{ronneberger2015unet}
O.~Ronneberger, P.~Fischer, and T.~Brox, ``U-Net: Convolutional Networks for Biomedical Image Segmentation,'' 2015. https://doi.org/10.48550/arXiv.1505.04597

\bibitem{Ostrom2023}
Ostrom, Q. T., Price, M., Neff, C., Cioffi, G., Waite, K. A., Kruchko, C., \& Barnholtz-Sloan, J. S. (2023).
CBTRUS statistical report: Primary brain and other central nervous system tumors diagnosed in the United States, 2016--2020.
\textit{Neuro-Oncology, 25}(Supplement\_2), iv1--iv95.
https://doi.org/10.1093/neuonc/noad149

\bibitem{bakas2017advancing}
S.~Bakas, H.~Akbari, A.~Sotiras, M.~Bilello, M.~Rozycki, J.~S.~Kirby, \emph{et al.},
``Advancing The Cancer Genome Atlas glioma MRI collections with expert segmentation labels and radiomic features,''
\emph{Scientific Data}, vol.~4, 170117, 2017.
https://doi.org/10.1038/sdata.2017.117

\bibitem{bakas2017gbm}
S.~Bakas, H.~Akbari, A.~Sotiras, M.~Bilello, M.~Rozycki, J.~S.~Kirby, \emph{et al.},
``Segmentation Labels and Radiomic Features for the Pre-operative Scans of the TCGA-GBM Collection,''
\emph{The Cancer Imaging Archive}, 2017.
https://doi.org/10.7937/K9/TCIA.2017.KLXWJJ1Q.

\bibitem{bakas2017lgg}
S.~Bakas, H.~Akbari, A.~Sotiras, M.~Bilello, M.~Rozycki, J.~S.~Kirby, \emph{et al.},
``Segmentation Labels and Radiomic Features for the Pre-operative Scans of the TCGA-LGG Collection,''
\emph{The Cancer Imaging Archive}, 2017.
https://doi.org/10.7937/K9/TCIA.2017.GJQ7R0EF.

\bibitem{sudre}
Sudre, C. H., Li, W., Vercauteren, T., Ourselin, S., and Jorge Cardoso, M. (2017). Generalised Dice Overlap as a Deep Learning Loss Function for Highly Unbalanced Segmentations. Deep learning in medical image analysis and multimodal learning for clinical decision support : Third International Workshop, DLMIA 2017, and 7th International Workshop, ML-CDS 2017, held in conjunction with MICCAI 2017 Quebec City, QC,..., 2017, 240–248. https://doi.org/10.1007/978-3-319-67558-9\_28

\bibitem{baid2021rsnabrats}
U.~Baid, \emph{et al.},
``The RSNA-ASNR-MICCAI BraTS 2021 Benchmark on Brain Tumor Segmentation and Radiogenomic Classification,'' arXiv:2107.02314, 2021.

\bibitem{WHO2022}
World Health Organization. (2022). "Cancer of the brain and central nervous system"
https://www.who.int/news-room/fact-sheets

\end{thebibliography}
\end{document}